\documentclass{iaus}
\usepackage{graphicx}

\title[PopII 1/2 stars: very high $^{14}$N and low
$^{16}$O yields] 
{PopII 1/2 stars: very high $^{14}$N and low
$^{16}$O yields}

\author[R. Hirschi]   
{R. Hirschi}

\affiliation{University of Basel, Klingelbergstr. 82, 
4056 Basel, Switzerland \break email: raphael.hirschi@unibas.ch}

\pubyear{2005}
\volume{228}  
\pagerange{1--7}
\date{?? and in revised form ??}
\jname{From Lithium to Uranium: Elemental Tracers of Early Cosmic Evolution}
\editors{V. Hill, P. Fran\c{c}ois \& F. Primas, eds.}
\begin{document}

\maketitle

\begin{abstract}
Nine 20$\,M_\odot$ models  were computed with metallicities 
ranging from solar, through $Z=10^{-5}$ ([Fe/H]$\sim$-3.1) 
down to $Z=10^{-8}$ ([Fe/H]$\sim$-6.1) and
with initial rotational velocities between 0 and 600\,km\,s$^{-1}$ to study
the impact of initial metallicity and rotational velocity \cite{H05}.
The very large amounts of $^{14}$N observed ($\sim$0.03$\,M_\odot$) are only produced at 
$Z=10^{-8}$ (PopII 1/2).
The strong dependence of the $^{14}$N yields on rotation and other parameters like the 
initial mass and metallicity may explain the large scatter in 
the observations of $^{14}$N abundance.
The metallicity trends are best reproduced by the models with 
$\upsilon_{ini}/\upsilon_c \sim 0.75$, which is slightly above the mean 
observed value for OB solar metallicity stars.
Indeed, in the model with $\upsilon_{ini}$=600\,km\,s$^{-1}$ at 
$Z=10^{-8}$, the $^{16}$O yield is reduced due to strong mixing.
This allows in particular to reproduce the upturn for C/O and a slightly decreasing [C/Fe],
which are observed below [Fe/H]$\sim$ -3.

\end{abstract}

\firstsection 
\section{Introduction}
Precise measurements
of abundances of low metallicity stars have recently been obtained
by \cite{FS5}, \cite{FS6} and \cite{IR04}. These provide new constraints for the
stellar evolution models (see \cite{CMB05}, \cite{Fr04} and \cite{Pr04}).
The most striking constraint is the need for primary $^{14}$N production in very low
metallicity massive stars. Rotation helps producing large amounts of 
primary $^{14}$N through mixing of newly synthesised carbon and oxygen during helium
burning (see \cite{MEM05} and contributions by these authors in this volume).
Other constraints are an upturn of the C/O ratio with a [C/Fe]
about constant or slightly decreasing (with increasing metallicity) at very low metallicities, 
which requires an increase (with increasing metallicity) of oxygen yields
below [Fe/H]$\sim$ -3. This seems hard to reproduce with rotating models were
mixing usually increases the size of the helium burning core
and therefore increases the oxygen yields. 

\section{Computer model \& calculations}
The computer model used here is the same as the one described in
 \cite{psn04}. Convective stability is determined 
by the Schwarzschild criterion. Overshooting is only considered for 
H-- and He--burning cores with an overshooting parameter, 
$\alpha_{\rm{over}}$, of 0.1 H$_{\rm{P}}$. 
Since the distribution of velocities at very low metallicities is not well
know,  
two initial rotational velocities were explored at very low metallicities. 
The first one is the same as at solar metallicity, 
300\,km\,s$^{-1}$.
The ratio $\upsilon_{\rm ini}/\upsilon_c$ decreases at very low
metallicities for the initial velocity of 300\,km\,s$^{-1}$
because stars are more compact at lower metallicities. 
However there are some indirect evidence (see \cite{CMB05} and \cite{MEM05} for details)
that stars rotate faster at low metallicities. Therefore, the 
second $\upsilon_{\rm ini}$ is 
500\,km\,s$^{-1}$ at Z=10$^{-5}$ ([Fe/H]$\sim $-3.1) and 600\,km\,s$^{-1}$
at Z=10$^{-8}$ ([Fe/H]$\sim$-6.1). These values are chosen such that the ratio of the initial velocity to the break--up velocity, 
$\upsilon_{\rm ini}/\upsilon_c$ ($\sim 0.75$), is slightly
larger than the mean observed value for OB solar metallicity stars (0.63).
The evolution of the models was followed until core Si--burning. 
The yields of these models were calculated in
the same way as in \cite{ywr05}. 

\section{Evolution of the structure}
The bulk of $^{14}$N is produced in the convective 
zone created by shell hydrogen burning. If this convective zone deepens enough
to engulf carbon (and oxygen) rich layers, then significant amounts of primary
$^{14}$N can be produced ($\sim$0.03$\,M_\odot$). 
This occurs in both the non--rotating model 
and the fast rotating model at $Z=10^{-8}$ but for different reasons.
In the non--rotating model, it occurs due to
structure rearrangements at the end of carbon burning similar to the third dredge--up.
In the model with $\upsilon_{\rm ini}=$600\,km\,s$^{-1}$ it occurs during shell helium burning 
because of the strong mixing of carbon and oxygen into the hydrogen shell burning zone.

The most interesting model is the one with
$\upsilon_{\rm ini}$=600\,km\,s$^{-1}$ at $Z=10^{-8}$. In the course of core helium burning, 
the increase in the
strength of the shell H--burning (due to mixing of C and O) is so important that the star 
expands and the 
convective He--burning core becomes and remains smaller than at the start of core
He--burning. The yield of $^{16}$O is closely
correlated with the size of the CO core. 
Therefore the yield of $^{16}$O 
is reduced due to the strong mixing 
(see also \cite{MEM05}). 

\section{Stellar yields}
The observational constraints at very low Z are a very
high primary $^{14}$N production, a flat or slightly decreasing [C/Fe] 
(with increasing metallicity) and an upturn in [C/O] for [Fe/H]$\lesssim$ -3 \cite{CMB05}.
This requires not only extremely high primary $^{14}$N production 
and a similar or larger $^{12}$C
production at very low Z but also a reduced $^{16}$O production in massive stars. 
All these criteria
are best fulfilled in the models with 
values of the ratio of the initial velocity to the break--up velocity 
($\upsilon_{\rm ini}/\upsilon_c \sim 0.75$) slightly
larger than the mean observed value for OB solar metallicity stars (0.63). It thus favours 
faster rotation at very low metallicities. 
The very large amounts of $^{14}$N observed 
($\sim$0.03$\,M_\odot$) are only produced at 
$Z=10^{-8}$ (PopII 1/2).
The strong dependence of the $^{14}$N yields on rotation and other 
stellar parameters like the initial metallicity and mass 
(see \cite{CL04} for the mass dependence at $Z=0$) may explain the large 
scatter in the observations of $^{14}$N abundance.

\end{document}